# Can Humans Really Discriminate 1 Trillion Odors?

Markus Meister
Caltech

## Abstract

A recent paper in a prominent science magazine claims to show that humans can discriminate at least 1 trillion odors. The authors reached that conclusion after performing just 260 comparisons of two smells, of which about half could be discriminated. Furthermore the paper claims that the human ability to discriminate smells vastly exceeds our abilities to discriminate colors or musical tones. Here I show that all these statements are wrong by astronomical factors. A reanalysis of the authors' experiments shows they are also consistent with humans discriminating just 10 odors. The paper's extravagant claims are based on errors of mathematical logic. Further analysis highlights the importance of establishing how many dimensions the perceptual odor space has. I review some arguments on the topic and propose experimental avenues towards an answer.

## Introduction

The present article begins with a critique and reinterpretation of a single publication (Bushdid et al., 2014), henceforth referred to as "The Paper". The Paper's title states that "Humans can discriminate one trillion odors." Thanks to the joint publicity efforts of *Science* magazine, Rockefeller University, and the Howard Hughes Medical Institute, the Paper's claims were broadly disseminated and eagerly taken up by the popular press. By now, many people "know" that humans discriminate a trillion odors, and that our color vision system pales miserably in comparison.

None of this is true. The authors were misled by failures in a mathematical method they designed. As a result, these claims have no basis. I will prove the assertion using three complementary approaches: First, I test the proposed analysis method on the human color vision system, for which we have enough mechanistic understanding to be confident in a simulation. The Paper's method says that humans can discriminate an infinite number of colored lights, whereas the correct number is ~1 million. Second, a reanalysis of the Paper's data on human smell discrimination shows that a simple model with just 10 discriminable odors is equally consistent with the observations. Finally, I track down the flaw in the Paper's logic.

For the non-expert reader it is worth emphasizing that the tests applied below are part of the standard repertoire of experimental science. The first concept is that of a "positive control experiment". Whenever a new assay or analysis are introduced, it is customary to ask for a positive control, namely a test of the method on a case where we already know the correct answer. Here the new assay includes the use of sphere-packing mathematics and arguments in high-dimensional spaces, which may easily fall out of the comfort zone of editors and reviewers. Thus a positive control is essential. Because the Paper makes explicit claims about human color vision, it is natural to ask whether the assay gives the correct answer in that system for the number of discriminable stimuli. That's what I try in Figure 2, and the Paper's method fails the test in dramatic fashion. The second concept involves "hypothesis testing". Generally speaking, one makes progress in science not by showing that data are consistent with one's favorite hypothesis, but when they are inconsistent with an alternate hypothesis. The Paper never tests an alternate to the "trillion odors" hypothesis against the data set. That's what I try in Figure 3, and the data are equally consistent with a trillion percepts and 10 percepts (and of course anything in between). In summary, I conclude that the Paper's method doesn't work, and the Paper's data set is uninformative.

## The Paper's approach

The reader is encouraged to read the original Paper. The literature cited there also serves as background to the present article, and I will add citations only as required. No special knowledge of the field of smell is needed to evaluate the Paper's analysis method, which is the focus of my critique. Here I give a brief summary of the authors' procedures.

1) The space of all possible odor stimuli is huge. There are likely several hundred thousand distinct chemicals that smell. Any mixture of those chemicals is a point in odor space. For practical purposes one will have to explore just a subspace of this giant diversity. If one determines the number of discriminable odors in that subspace, that will be a lower bound on the total number of discriminable odors.

2) The Paper's experiments were limited to a subspace spanned by 128 substances. From these primary odors the authors made mixtures by drawing a number ($N$) of primary odors and mixing those in equal parts. In other words, each primary odor is either present or absent in the mixture.

3) In this space some mixtures are very similar to each other, for example if they share 29 of the 30 primary components and differ in only 1. Others are very different, for example when they share none of the primary components. For any two odors in the space the number of unshared components can be defined as their "distance". The authors assume that the ability of humans to discriminate two odors improves systematically with this distance.

4) Now the goal is to determine the critical distance at which two odors become discriminable. The authors probe this systematically by making pairs of mixtures with $N$ components of which $M$ are unshared, and testing those for discrimination by human subjects. Indeed they find that the probability of discrimination increases with $M$ (Paper Fig 3B). They define the critical distance $D$ as that separation $M$ at which 50% of the mixture pairs are discriminable.

5) With these assumptions, two odors closer than $D$ tend to smell the same, whereas two odors separated by more than $D$ will smell different. To determine how many discriminable odors there are, the authors ask how many regions of diameter $D$ can fit in the original 128-dimensional odor space.

6) This is analogous to the problem of packing spheres in high-dimensional spaces and the authors compute the number of packable spheres of diameter $D$ by methods of combinatorics (Paper Fig 3D). They put that number in the title of the Paper.

### Failure on a simple 3-percept model

To illustrate failure of this analysis I first apply it to a deliberately simple toy model of an olfactory system. Imagine a bacterium that has only 3 odor percepts: “yum”, “yuck”, and “meh”. In the simulation, we choose 128 primary odors and assign them randomly to be either attractant (sensory input +1) or repellent (-1). We make mixtures of 30 odors from the set of 128. The bacterium responds to a mixture by simply adding the sensory input from the component odors. If the sum is less than -2 it says "yuck"; if it is greater than +2 it says "yum"; and from -2 to +2 it says "meh". Two odor mixtures are considered discriminable if the bacterium responds differently to them. Following the Paper’s procedure we make odor mixtures that share a certain number of odors and plot the fraction of discriminable mixtures vs the odor overlap (Fig 1a, compare to Paper Fig 3B). The critical value of 50% discriminability is reached when the mixtures share 10 of the odors. This leads to an estimate of ~$3 \cdot 10^8$ discriminable stimuli (Fig 1b-c, see Paper Fig 3D). So by the standards of the Paper, this bacterium can discriminate more than 100 million odors. Yet we know by construction that it has only 3 odor percepts.

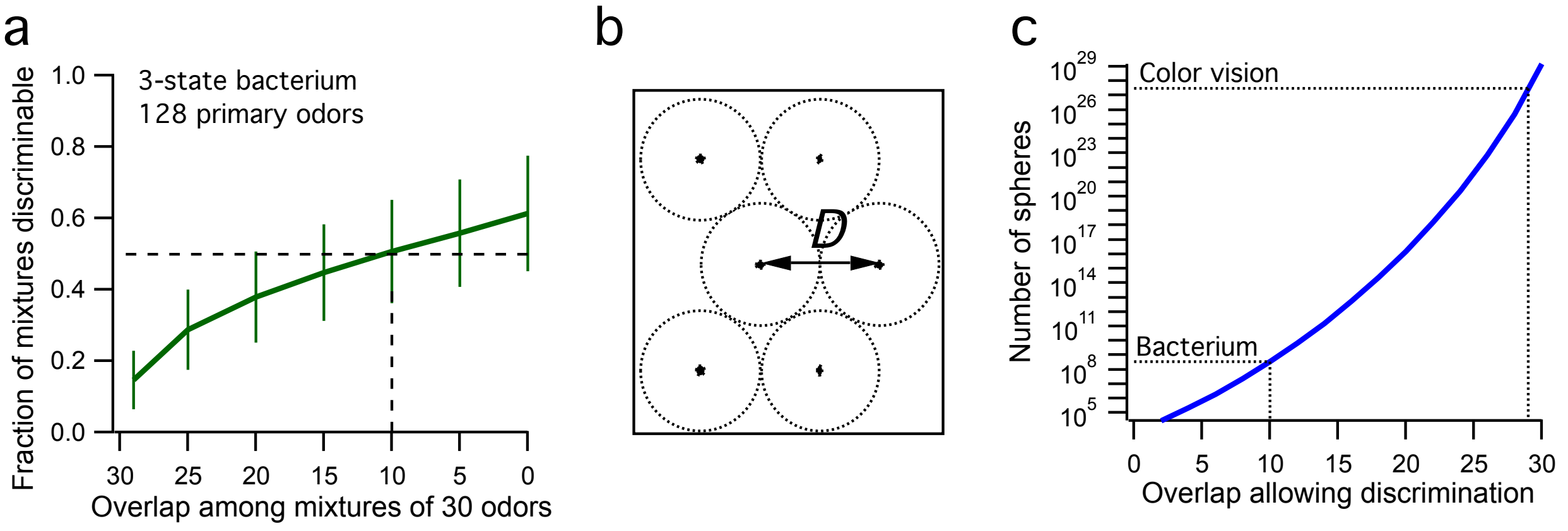

**Figure 1:** Model of olfaction in a bacterium. **a:** This 3-state bacterium simply counts how many odors in the mixture are attractants vs repellents, and classifies the result into 3 percepts (see text). Two odor mixtures are discriminable if they cause different percepts. The simulation followed the Paper’s procedures, yielding the fraction of discriminable mixtures as a function of the number of odors, $O$, that they share. Mean ± SD over 1000 repeats using different random assignments of the primary odors. Horizontal dashes: criterion for critical distance (50% discriminable pairs). Vertical dashes: critical distance $D$=30-$O$=20. **b:** Points in odor space separated by a distance $D$ cause different odor percepts. Counting how many such points exist in the space is like trying to pack spheres of diameter $D$ to fill the space as efficiently as possible. **c:** The number of such spheres in 128-dimensional space as a function of the discriminable overlap $O$ among 30-odor mixtures, computed as in the Paper. The value $O$=10 from panel (a) yields ~$3 \times 10^8$ spheres.

One gets the sense that there is something amiss with the Paper’s procedure. Yet the toy model differs from the human odor tests in several ways. The toy bacterium actually has a name for each odor in the space (one of 3 possible names) whereas humans were not asked to name the odor. Furthermore, for any given mixture pair the bacterium's response is deterministic, whereas humans varied in their response. Thus it would be useful to test the analysis on a system in which one can exactly replicate the human psychophysics methods used in the Paper's odor discrimination tests.

### Failure on the color vision system

A prominent claim in the Paper is that olfaction vastly outperforms color vision in terms of discriminable stimuli. In making this comparison, the authors compare their own estimate of a trillion distinct olfactory percepts to the literature's estimates of 1 million color percepts. However, studies of color vision used a very different procedure to determine the number of discriminable lights (more on this below). Here I ask what number the Paper's methods would produce. Fortunately we know enough about the rules and mechanisms of color perception to simulate this with great confidence, so there is no need to actually perform new color discrimination trials.

The space of all colored lights has infinite dimensions. Each light can be characterized by its wavelength spectrum $S(\boldsymbol{\lambda})$, which specifies how much power exists at each wavelength $\boldsymbol{\lambda}$. Since the wavelength can take on any positive real value, the $S(\boldsymbol{\lambda})$ are functions of a continuous variable, and thus have infinite dimensions. However, the perceptual space of human color vision has only 3 dimensions (Wandell, 1995). One way to express this is that all perceivable colors can be produced by a graded mixture of just 3 primary lights. The reasons for this are well known: The human retina has 3 types of cone photoreceptor (red, green, and blue), and the perceived color depends on the excitation of those 3 cone types. Furthermore, the excitation of each cone depends only on the rate of photon absorption by its visual pigment. In turn, that rate is a linear function of the spectrum of the light, namely the projection of the light spectrum onto the absorption spectrum of the pigment. If two lights are mixed together, the resulting excitation is the sum of the effects of the individual lights. Therefore human color vision begins by projecting the infinite space of lights down onto a subspace of just 3 dimensions, spanned by the 3 cone excitations. This subspace has been probed extensively in psychophysical experiments. Typically the subject is shown two lights side-by-side and asked whether they appear different. By systematic probing of the 3-dimensional space one finds there are upward of 1 million discriminable lights, in the sense that any two of them will look different in a pairwise comparison (Masaoka et al., 2013).

For the purpose of simulation, I will therefore consider a perceptual space with 3 axes: $R$, $G$, and $B$ (Fig 2a). For simplicity, the 3 variables will range from 0 (dark) to 1 (bright), so the perceptual space is a unit cube. Any given light stimulus is represented by a vector in the unit cube. The color vision system adds some noise to that vector along each of the 3 dimensions. As a result, two lights are discriminable if their vectors are separated by more than the noise. The level of noise is chosen so that a difference of 0.01 along any dimension is discriminable, which gives 1 million discriminable vectors in the unit cube.

Now one can implement the Paper's procedures: Choose at random 128 primary lights in this space and make them of equal intensity. Then produce mixtures of 30 lights from those primaries; again design different classes of mixtures that vary in the number of shared components. Within each mixture class, present pairs of stimuli to the model and ask whether they can be discriminated, following the “odd-man-out” procedure. Plot the fraction of discriminable pairs against the mixture overlap. Find the overlap at which that fraction is 50%, and take the number of unshared components as the critical distance *D* for discrimination. Compute how many spheres of size *D* fit in the original 128-dimensional space.

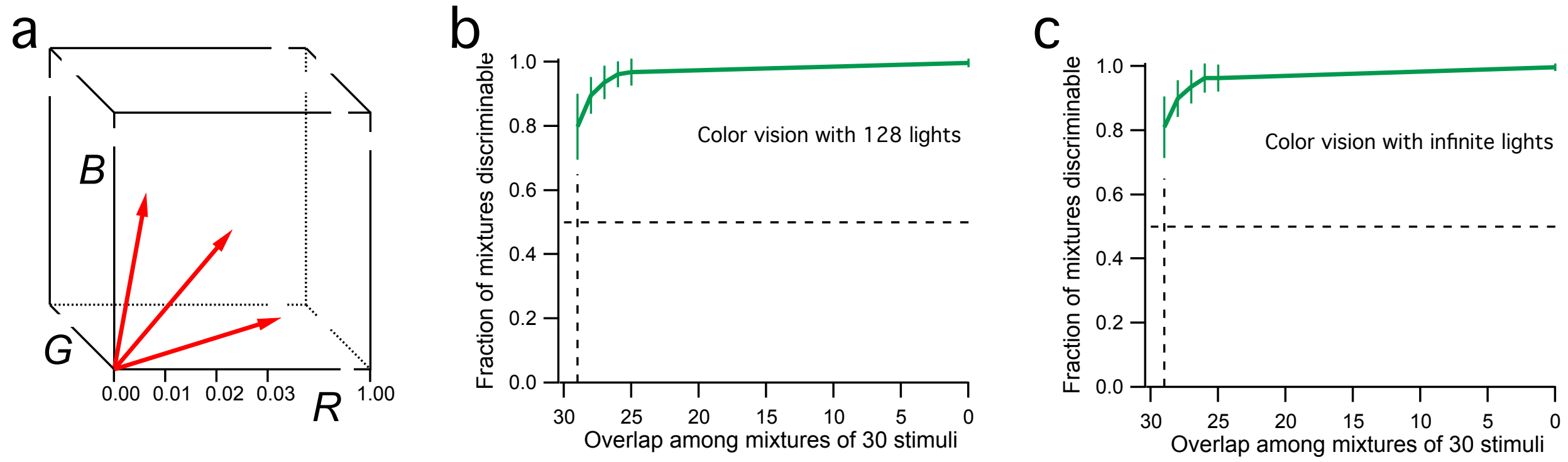

**Figure 2:** Model of human color vision. **a:** The RGB color cube with 3 of the 128 primary colors represented by vectors from the origin. Tick marks represent just noticeable differences, e.g. along the *R*-axis. **b:** The fraction of discriminable 30-light mixtures as a function of their overlap. Mean ± SD over 1000 repeats using different random assignments of the 128 lights. 30 lights per mixture, 20 mixture pairs per class, 26 subjects per pair. Horizontal dashes: criterion for critical distance. Vertical dashes: critical distance. **c:** As in panel b but with the mixture components drawn at random from all possible directions in the space rather than from a preselected set of 128 primaries. The results are almost identical.

From this simulation, I find that all 30-light mixtures are discriminable at >50% performance if they differ by as little as one component (Fig 2b). This is to be expected. Each of the primary lights has a vector length of 1/30 (see Methods). This means that a single unshared component can separate the two mixture vectors by more than 1/30 in the RGB space. But a separation of just 1/100 along any axis is discriminable. So the critical distance *D*=1. From this one calculates there are more than $10^{28}$ discriminable colors in the 128-dimensional space probed here (Fig 1c, Paper Figs 3D, S1B).

Actually, the number of discriminable colors by this argument is much higher, even infinite. To drop down to the 50% discriminability criterion used by the authors in defining *D*, we have to make much larger mixtures of ~60 lights that differ by just one component. The number of possible mixtures of that kind is about $10^{37}$. Furthermore, there is no reason to limit the starting space to just 128 primary lights. There is an infinity of spectra that are physically achievable lights, so we could have started with a subspace of arbitrarily high dimensionality. One can simulate infinite dimensionality by choosing for each of the mixtures a different random set of 30 normalized vectors from the cube (again maintaining the specified overlap among mixture pairs). As shown in Fig 2c, the critical distance under those conditions is still 1. Thus the Paper’s methods would assert that humans can discriminate an infinite number of colors. That is much bigger than 1 trillion, so color vision wins over smell, at least by the internal logic of the Paper.

## Humans can discriminate at least 10 odors

From this failed positive control it is clear that the analysis can vastly overestimate the number of discriminable stimuli. Therefore one wonders whether the human discrimination data reported in the Paper could be explained by a much smaller number of percepts. That is indeed the case.

I consider a simple model of odor processing that can simulate the human psychophysics experiments in the Paper (Fig 3a): Suppose the olfactory system projects all odors onto a single dimension, and let us use the vectors on the unit circle for that. So the 128 primary odors map onto 128 unit vectors with random angles. We suppose that a mixture of odors gets mapped into the sum vector of all the components, normalized again to unit length. Finally, the angle of this vector gets corrupted by some perceptual noise. Two vectors around the circle will be discriminable if they are separated by more than the noise.

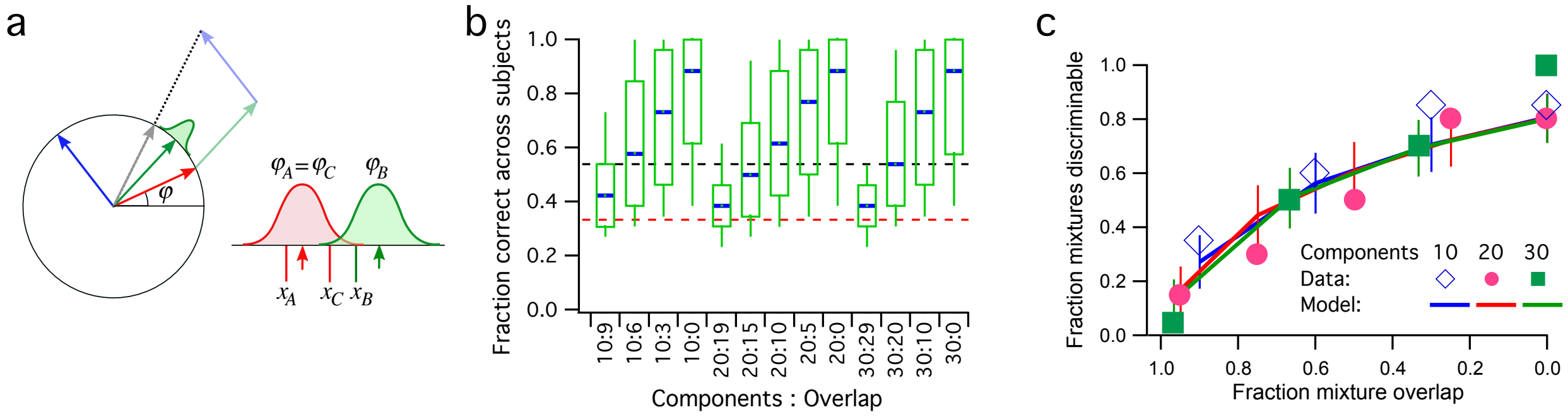

**Figure 3:** A simple simulation of the human smell experiments. **a, left:** Each primary odor gets mapped into a unit vector (e.g. red, green, blue). Mixtures of odors get mapped into the normalized sum vector (gray). **right:** When a subject sniffs an odor vial, the odor angle is corrupted by Gaussian noise and a value is drawn from that distribution. Here 3 vials were presented, two (A and C) containing the identical odor and a third (B) a different odor. This produced response variables $x_A$, $x_B$, and $x_C$. On this trial, $x_B$ and $x_C$ are closest to each other, so the subject (incorrectly) identifies A as the odd odor. **b:** Discriminability of odor mixtures under this model (compare to Fig 2C of the Paper). Mixtures were simulated according to the Paper's procedure with 10, 20, or 30 components and varying overlap. In each mixture class, 20 mixture pairs were made from the 128 primary odors. Each mixture pair was presented to 26 subjects (taken to be identical), and the fraction of correct identification determined across subjects. Box-and-whisker plot shows the distribution of that fraction with percentiles 10, 25, 50, 75, 90. Average over 1000 repeats of the procedure using different random numbers. Red dashes: chance performance. Black dashes: criterion for discriminability (14/26 correct). **c:** Fraction of discriminable mixtures as a function of their overlap (compare to Fig 3D of the Paper). This is the fraction of mixture pairs in each class that exceeds 50% correct identification across subjects (above the black line in panel b). Lines are mean ± SD from the simulation, over 1000 repeats with different random numbers. Symbols are data from the Paper. The model used Gaussian noise with a SD of 0.4 radians.

This encoding model can simulate the human subject experiments in detail: As in the Paper, for every "odd-man-out" trial, 3 mixtures are presented to the model, of which two are identical and the third is a different target mixture. The model maps all 3 onto the unit circle, adds noise to each, and asks which two are closest to each other. Then it reports the more distant one as the odd odor. If that corresponds to the target odor, it is a "correct" decision (Fig 3a). This procedure was run through all the same paces used in the Paper: 26 human subjects x 20 odor mixture pairs x 13 categories of overlap. Qualitatively, mixtures that share a large fraction of components produce sum vectors at nearby angles, which makes them less discriminable because of the perceptual noise (Fig 3b, compare to Paper Fig 2C). Again one can compute the fraction of discriminable odor mixture pairs, namely those correctly identified in >50% of all trials. In Fig 3c, I plot the results of the simulation along with the Paper's data (from Paper Fig 3B).

This model has a single parameter, namely the amount of perceptual noise. Conveniently that is also the only number we want to know. Nothing else is adjustable. With a noise value of 0.4 (SD of the Gaussian noise measured in radians) the fit to the data is quite good (Fig 3c). Almost all the data points are within the 1 SD error bars. At noise = 0.3 the model does somewhat better than the humans, at noise = 0.5 somewhat worse.

For a noise value of 0.4, how many mutually discriminable stimuli are there? We should use the Paper's standard for discriminability of an odor pair: 50% correct identification in the "odd-man-out" odor test. With that criterion, one can place at most 10 vectors around the unit circle such that each is discriminable from its neighbors. Therefore, the published measurements are equally consistent with a model in which humans can distinguish just 10 olfactory stimuli from each other.

Many of us have experienced more than 10 smells, so this study did not come close to exploring the richness of human olfactory experience. Why is the lower bound from this odor psychophysics study so weak? Actually, the result is just about expected from the effort expended. To confirm that 10 odors are mutually discriminable by brute force one has to compare each odor to every other one. The authors did 260 pairwise comparisons, of which only about half were discriminable. So one expects to find evidence for about $\sqrt{130}$ distinct odors, close to what was obtained.

The ring model of Fig 3 assumes that odor percepts lie in a 1-dimensional space. If one allows for higher numbers of dimensions, then the predicted number of discriminable odors increases, approximately exponentially with the dimensionality of the perceptual space (L. Abbott, E. Schaffer, & R. Axel, personal communication). Since we do not know the dimensionality of odor space, the Paper's results are equally consistent with 10 or a trillion discriminable odors or anything in between.

**Where is the flaw?**

The failure of the method occurs after the sphere-packing estimate (Fig 1b). It involves a step that is never mentioned but implicit in the Paper's procedure: the assignment of odor percepts to the different spheres. There are at least two problems that can lead to enormous overestimates of the number of distinct odor percepts:

1) The authors assume that every one of the spheres packed into the space corresponds to a different odor percept. But this is unwarranted. From the sphere-packing argument we can only conclude that *neighboring* spheres correspond to different percepts. Thus the same percept may recur over and over again within the odor space.

The situation can be understood already in 2 dimensions. Fig 4a shows a set of close-packed pennies on the desktop. Every penny is different in color from its neighbors. More importantly every penny in the whole space has a different color from every other penny. If the pennies correspond to odor percepts one could legitimately say that this organism discriminates as many odors as there are pennies. However, there are many other ways to color close-packed pennies so that no adjacent ones have the same color. In fact, 3 colors are sufficient (Fig 4b), so just 3 percepts could account for an infinite number of spheres in a 2D odor space.

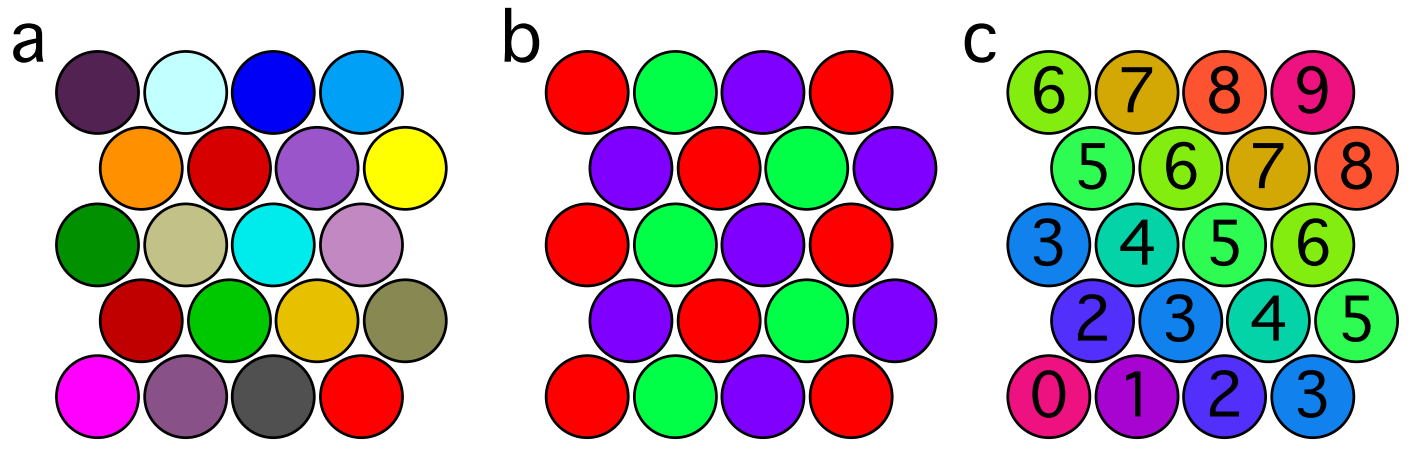

**Figure 4:** Coloring close-packed spheres in 2 dimensions so no nearest neighbors have the same color. **a:** All colors different. **b:** Three colors suffice. **c:** Progression of colors in one dimension only.

One might object that this is an esoteric arrangement of the 3 percepts. What kind of olfactory system would produce this strange periodic recurrence of the same percept? However, there is a perfectly natural color progression that also provides enormous savings of colors. In the arrangement of Fig 4c the percept stays the same along one dimension of the space and varies along the orthogonal dimension, yet all neighboring spheres are distinct. So the number of pennies one can pack this way goes as the square of the number of colors available. In 128-dimensional space, the number of hyperspheres one can pack this way will go as the 128$^{th}$ power of the number of percepts.

So the key failure point of the Paper's logic is to assume that all the close-packed hyperspheres are labeled with different odor percepts. One can trace this back to the unstated assumption that the odors should systematically become more discriminable at greater distances in this space, *and equally in every direction*. Here I showed that there is a simple and natural way to violate that assumption, namely if the percept depends only on a single function of the coordinates. Then the odors become more discriminable with distance along one dimension, but remain indiscriminable along all other dimensions. More generally, if the percepts live in some low-dimensional space (e.g. 3D for color, 2D for pure tones) and then one embeds that space in 128 dimensions, this will produce a similarly efficient labeling of the close-packed spheres. This is what causes the astronomical overestimates in the Paper's analysis when applied in the three model simulations above.

2) The above arguments cannot fully explain the results obtained with the model of the 3-state bacterium. Here there are only 3 possible colors. Yet, in 128-dimensional space every sphere has at least 256 nearest neighbors. Clearly it is impossible to distinguish a sphere from all its nearest neighbors with 3 colors. This brings us to a second weakness of the method: the definition of the critical distance *D*. This is taken as the distance at which 50% of odor mixture pairs are discriminable. In other words, two points separated by *D* in odor space will produce a different percept 50% of the time. So when assigning percepts to the spheres in 128-dimensional space, we merely need to ensure that each sphere is distinct from 50% of its neighbors. That can be done trivially with just 2 colors, painting alternating spheres white and black. That makes every sphere different from 50% of all its neighbors. The 3-state bacterium can do a bit better, which is why discriminability rises to almost 2/3 at large distances (Fig 1a).

Would it help to raise the criterion for discriminability to a higher percentage of mixture pairs? If one raises it to 90% then 10 odor percepts are sufficient to color the entire space of spheres. A popular estimate for the number of discriminable odors (though without solid scientific basis, as reviewed in Bushdid et al., 2014) is around 10,000. To ensure that a sphere-coloring system uses at least 10,000 odors, one would need to raise the criterion for discriminability in the definition of *D* to 99.99%. So there needs to be a data point in Paper Fig 3B with an ordinate of 99.99%. By the methods of the Paper, that requires human subjects to perform several tens of thousands of pairwise comparisons in just one mixture class. This would entail an extraordinary experimental effort.

**How to proceed?**

The arguments above illustrate that the procedure for exploring sensory spaces advocated in the Paper does not work. Fundamentally the method presumes that the space of odor percepts has at least 128 dimensions. Furthermore, the 128 primary odors chosen must represent “orthogonal” directions in that space, so that the percept varies with distance in the same manner and independently along each of these dimensions. Nothing in the Paper suggests why we should believe this, and the assumptions seem implausible *a priori*. Violation of those assumptions, for example if the true dimensionality of odor space is less than 128, leads to dramatic failure of the estimate. How can one do better?

To get a satisfying answer to the question in the title, we want to know what is the largest set of stimuli such that every stimulus can be discriminated *from every other one*, not just from nearest neighbors. Classic studies of color vision and tone hearing have achieved that. For color vision, the answer is approximately $n$ = 1 million. How can one come to that conclusion without doing $n^2$ discrimination tests? This is possible only by taking advantage of the systematic structure of the perceptual space. For example, if light *A* looks more red than light *B* and *B* looks more red than *C*, then one can trust that *A* will look more red than *C*. The same applies in tone hearing: If tone *A* sounds lower than tone *B*, and *B* sounds lower than *C*, then *A* will sound lower than *C*. This transitivity of discrimination means that it is sufficient to measure discrimination of neighboring stimuli to ensure mutual discrimination of all stimuli in the set, which reduces the experimental burden from order($n^2$) down to order($n$). Effectively these studies work in a low-dimensional stimulus space (R, G, B for lights; frequency and amplitude for pure tones), within which the progression of the percepts is monotonic, such that transitivity applies. So success of these percept-counting studies relied entirely on recognizing early on that the perceptual space is low-dimensional.

## The dimensionality of odor space

How to apply those insights to olfaction? Clearly one needs to first answer "how many dimensions span the perceptual space of odors?" The number of 400 human odorant receptor types certainly sets an upper bound on the dimensionality of the perceptual space. And one commonly hears the argument: "Why would Nature make so many receptors, unless it is to discriminate all possible patterns of activation?" However, there may be another explanation: We need that many receptors simply to *sense* the molecules of interest, not to disambiguate all possible mixtures of those molecules.

The bacterium E. coli illustrates this idea: It has 5 different chemoreceptor proteins with different ligand spectra. In principle, E. coli could therefore analyze odor mixtures in 5-dimensional space and react differently to every possible mixture. But it doesn't. The outputs of all 5 receptors readily converge on a single variable, namely the concentration of the CheY signaling molecule (Grebe and Stock, 1998). So E. coli projects the 5-dimensional receptor space onto a 1-dimensional perceptual space of attraction/repulsion. Why then does it need 5 receptors, including different receptor proteins for aspartate and for serine, both amino acids? Presumably it is difficult make a generic amino acid receptor with high sensitivity. Both the amino and the carboxyl ends of the molecule vary in their charge distribution depending on pH, and it would be tough to design a binding pocket with sufficient affinity to work under all conditions. So instead the two receptor proteins focus on other more stable portions of the ligand, but those are also unique between serine and aspartate.

One can apply the same idea to odorants in the human nose. The molecules of interest are there at micromolar concentrations or less, in a mucus soup of other components at millimolar concentration. To sense the odorants separately from the mucus, a receptor needs to bind them with high affinity. That means many contact sites between binding pocket and ligand, which in turn leads to selectivity for the shape of the ligand. Even if the olfactory system just wanted to distinguish odors along one dimension (attractive/repulsive), it would be impossible to make a receptor that is selective just for the attractants. Instead Nature makes many receptors that are each selective for small groups of related molecules and then sorts their signals apart using the nervous system.

In this picture the dimensionality of receptor space is determined by molecular principles involving the number of ligands of interest, their relevant concentrations, the energetics of ligand binding, and the design limitations of protein structures. The dimensionality of perceptual space, on the other hand, is governed by behavioral and ecological constraints: the nature of olfactory cues in the environment, the kinds of decisions the animal makes based on odorants, and the need to associate new odors with unusual events. There is no principled reason that this perceptual space should have the same dimensionality as the receptor space. And we have the neural circuits of the olfactory system to create an arbitrary map from one space onto the other.

Some psychophysical studies suggest now that the effective number of dimensions may be quite low, and that the dominant axes of the odor space are related systematically to the physical characteristics of odorous molecules (Secundo et al., 2014). Another relevant observation is that mixtures containing many (>20) diverse odorants tend to smell alike, even if they don't share any molecular components (Weiss et al., 2012), a phenomenon that has been termed “olfactory white”, in analogy to the mix of many colored lights. This suggests that the dimensionality may be around 20.

A useful experimental approach might then be to rigorously measure the dimensionality of perceptual space *at least at one point*. For example, choose $N$ primary odors, and consider arbitrary mixtures of those as the odor space. Define the “white point” as the mixture of all those odors at half concentration, i.e. the odor vector $\mathbf{w}=(0.5,\ldots,0.5)$. How does odor perception vary as the stimulus deviates a little from this point? To first order, the discriminability $d$ between the white odor at $\mathbf{w}$ and an odor at $\mathbf{w}+\mathbf{x}$ will vary as a quadratic form of the deviation vector $\mathbf{x}$, namely: $d = \mathbf{x}^{T}\ \mathbf{S}\ \mathbf{x}$. We want to know the sensitivity matrix $\mathbf{S}$. By definition it is positive definite and symmetric, and thus has $N(N+1)/2$ unknown components. So one needs to measure the just-discriminable-distance along $N(N+1)/2$ directions from the white point. Clearly this is an experimental challenge, but it seems plausible at least for $N$=20. If so, then the structure of the matrix $\mathbf{S}$ can reveal the dimensionality: In particular, if it has just a few large eigenvalues, those identify the relevant directions in odor space. By contrast, if all eigenvalues are comparable, then the perceptual space has dimensions higher than $N$.

Animal studies can play an important role here. Mice are readily trained to distinguish odors, even closely related mixtures. More importantly, they offer an opportunity to stimulate the receptor neurons directly, by optogenetic activation of the olfactory bulb (Spors et al., 2012). In a suitably engineered animal one could drive arbitrary activation patterns of the different olfactory receptor types by shining patterned light onto the glomeruli in the olfactory bulb. This approach promises several benefits in a study of odor dimensions: First, it does away with the tedium of olfactory stimulation, such as mixing dozens of vapors, switching valves, flushing tubes, and waiting for odors to dissipate. Using light, a different combination of receptors can be driven with millisecond precision and at high repetition rates. Also, this method allows patterns of stimulation that may not ever occur with natural odorants; one could then test if the perceptual space is shaped to the ecology of real odors.

In a way, one can think of the olfactory bulb surface as a retina for the smell system. Olfactory objects are represented by spatio-temporal patterns on this surface, and the downstream neural circuits are busy identifying, discriminating, or learning those spatio-temporal patterns. The optogenetic approach simply takes the analogy one step further by using light as a stimulus. At the same time, there is a parallel effort ongoing in vision science to determine the dimensionality of human pattern vision. It is clear already that the number of dimensions is much lower than the number of pixels on the retina. One can make pairs of visual images that have very different effects on the retina, but look the same to human subjects (Freeman and Simoncelli, 2011). And a systematic approach to measuring dimensionality of pattern vision is beginning to yield results (J. D. Victor, personal communication). Based on these developments, I suggest that pursuing the analogy between smell and pattern vision will be much more fruitful than the analogy to color vision.

Regardless of approach though, determining the dimensionality of the space of odor percepts is a precondition to estimating the number of distinct percepts. The recognition that color space is three-dimensional has had enormous impact in science, art, and technology, as anyone reading this on a color monitor will confirm. The search for a similar basis set for odors has fascinated scientists, engineers, and perfumers for some time (Gilbert, 2008). Even proving that a low-dimensional basis does not exist would be a major advance.

## Cited Literature

## Methods

All simulations and graphics were produced with Igor (Wavemetrics). Annotated code is available from the author. Simulation of the three models (Figs 1-3) followed the same process as the human odor tests performed in the Paper:

1. Selection of the 128 primary stimuli. These were drawn at random from the stimulus space. Fig 1: binary distribution over {-1,+1}. Fig 2: Uniform distribution of R, G, and B over [0,1], followed by normalization to a length of 1/30. Fig 3: Uniform distribution on the unit circle. These are conservative choices, in that a random set of primaries does not cover the stimulus space particularly well, and thus will produce mixtures that occupy only a portion of the space. This will therefore underestimate the number of discernible stimuli. By contrast the Paper chose odors that were "well distributed in both perceptual and physicochemical stimulus space."
2. Creation of mixtures. Pairs of mixtures containing $N$ primaries of which $O$ are shared were created by choosing at random $2N-O$ components from the set of primaries, adding the first $O$ to both mixtures, the next $N-O$ to mixture 1 and the last $N-O$ to mixture 2. For every class of mixtures (i.e. combination of $N$ and $O$), 20 mixture pairs were created. The rules for combining the stimulus values in a mixture were as follows: Fig 1: Simple addition of the binary stimulus values. Fig 2: Addition of the primary vectors. Fig 3: Addition of the unit vectors followed by normalization to unit length. This normalization emulates the elimination of odor intensity cues in the Paper.

3. Discrimination test. Every pair of mixtures was presented to the model to determine whether it was discriminable or not. Fig 1: The model classifies every mixture of *N*=30 odors into the percepts “yum”, “meh”, or “yuck” depending on whether the sum of stimulus values is >2, from 2 to -2, or <-2 respectively. Two mixtures that fall into different percepts are discriminable. Fig 2: The model performs an odd-man-out discrimination test among 3 samples as done for human subjects in the Paper. Two of the samples contain mixture 1 and the third contains mixture 2. Perceptual noise was simulated by adding a Gaussian random variable to each of the 3 coordinates of the mixture vectors. The resulting 3 sample vectors are inspected and the two with the smallest distance are declared to be the same. For every mixture, this is performed 26 times, with different draws from the noise distribution. As in the Paper, a mixture pair is declared discriminable if the model gives the correct response on >50% (14 or more) of those trials (note chance performance is 1/3). In the Paper, these 26 trials were performed by 26 different human subjects. There was some indication that different subjects had different abilities, but the analysis merged them all. In my simulation, all trials are done with the same amount of perceptual noise. The noise magnitude was chosen so that stimuli separated by a distance of 0.01 are just discriminable. This leads to ~1 million distinct percepts in the RGB stimulus space, a conservative choice, because the empirical estimates of that number are somewhat larger. Fig 3: This followed the same odd-man-out procedure as for Fig 2. Perceptual noise was simulated by adding a Gaussian random variable to the angle of the mixture vector (Fig 3a).
4. Discriminability of mixture classes. For each class of mixtures I computed what fraction of the 20 were discriminable (by the criteria in 3), and plotted this against the mixture overlap (Figs 1a, 2b-c, 3c). To estimate the reliability of the simulation, the entire procedure was repeated 1000 times (with different random numbers) and the plots show the mean and standard deviation of the outcome. Fig 3c shows my simulation along with the data from the Paper’s human experiments.
5. Estimating the number of discriminable stimuli. From the graph of discriminability vs overlap, the critical distance *D* was taken to be the number of unshared stimuli that allows 50% discriminability in that class of mixtures. The number of regions of diameter *D* that can fit into the stimulus space was computed using the formula derived in the Paper (Fig 1c). If there are *C* primaries total and *N* primaries per mixture, then the number *S* of such regions is claimed to be

$$S = \frac{\binom{C}{N}}{\sum_{R=0}^{D/2} \binom{N}{R} \binom{C-N}{R}} .$$

**Acknowledgements**

Thanks to Adam Shai for extended discussions.